\documentclass[11pt]{article}

\advance \textheight by \topskip
\makeatletter
\@addtoreset{equation}{section}
    
\makeatother

\def\vx{\vec{x}}

\def\vA{\vec{A}}

\def\vp{\vec{p}}
\def\vv{\vec{v}}
\def\vP{\vec{P}}

\def\cP{{\cal{P}}}
\def\cX{{\cal{X}}}

\def\cT{{\cal{T}}}
\newcommand{\p}{\partial}

\def\Di{\displaystyle}

\usepackage{amsmath,amssymb}

\begin{document}

\title{
{\bf Nonrelativistic anyons in external electromagnetic
field
}}

\author{
  {\sf P.~A.~Horv\'athy${}^a$}\footnote{Parc de Grandmont,
 F-37200 TOURS (France). e-mail: horvathy@univ-tours.fr}
  \
  {\sf and M. S. Plyushchay${}^b$}\footnote{e-mail:
  mplyushc@lauca.usach.cl}\\
  \\[4pt]
{\small \it ${}^a$Laboratoire de Math\'ematiques et de
Physique
Th\'eorique}\\
 {\small \it Universit\'e de Tours, France}\\
 {\small \it ${}^b$Departamento de F\'{\i}sica,
Universidad de Santiago de Chile}\\
{\small \it Casilla 307, Santiago 2, Chile}
 }
\date{}

\maketitle

\begin{abstract}

The first-order, infinite-component field equations
we proposed before
for non-relativistic anyons (identified with
particles in the plane with noncommuting coordinates)
are generalized to accommodate arbitrary
background electromagnetic fields.
Consistent coupling of the underlying classical system
to arbitrary fields is introduced; at a critical
value of the magnetic field, the particle follows a
Hall-like law of motion.
The corresponding quantized system
reveals a hidden nonlocality if  the magnetic
field is inhomogeneous.
In the quantum Landau problem
spectral as well as state structure (finite vs. infinite)
asymmetry is found.
 The bound and scattering states, separated by the
critical magnetic field phase,
behave as further, distinct phases.
\end{abstract}

\vskip.5cm\noindent

\section{Introduction}

In a previous paper \cite{HP2} we proposed  an infinite set
of
first-order 
 field equations for non-relativistic anyons, namely
\begin{equation}\left\{\begin{array}{ll}
i\p_{t}\phi_{k}+\displaystyle\sqrt{\frac{k+1}{2\theta}}\,
	\displaystyle\frac{p_{+}}{m}\,\phi_{k+1}=0,
	\\[16pt]
	p_{-}\,\phi_{k}
	+\displaystyle\sqrt{\frac{2(k+1)}{\theta}}\,
	\phi_{k+1}=0,
    \end{array}\right.
     \label{infLL}
\end{equation}
k=0,1,\dots.
Here $p_\pm=p_1\pm ip_2$, and we assume for definiteness
that the non-commutative
parameter $\theta$ (whose physical dimension is $m^{-2}$,
see below) is positive.
Grouping the upper and  lower equations, respectively,
(\ref{infLL})  can also be presented in the form
\begin{equation}
     \left\{\begin{array}{c}
D|\phi\rangle=0
\\[10pt]
\lambda_-|\phi\rangle=0
\end{array}\right.,
\qquad
\begin{array}{cc}
D=i\p_t-h,\hfill
&h=\vec{p}\cdot\vec{v}-\frac{1}{2}m\, v_+v_-,
\\[10pt]
\lambda_-=p_--mv_- ,\quad&
\end{array}
\label{DL}
\end{equation}
where
$|\phi\rangle=\sum_{k=0}^\infty\phi_{k}|k\rangle_v$. The
$|k\rangle_v$ are the Fock states of the velocity operators
$v_{\pm}=v_1\pm iv_2$, $v_-|0\rangle_v=0$.
The latters are analogous to the  $\alpha$
matrices of Dirac,
but are associated with an infinite-component, Majorana -
type representation of the planar Galilei (rather then
Lorentz) group.

The first of these equations is reminiscent of the usual
Dirac equation (more precisely, of its non-relativistic
counterpart due to L\'evy-Leblond \cite{LL})
in that it is of the first order in the derivatives.
$h$ is the Hamiltonian.

$\lambda_-$ measures the
difference between the canonical
($p_i$) and the mechanical ($mv_i$) momenta.
The second equation can be viewed as a constraint
$\lambda_-\vert\phi\!>_{phys}=0$ which specifies the
physical subspace as composed of coherent states of the
velocity.
This  makes the spectrum bounded from below
by freezing the internal spin degrees of freedom \cite{HP2}.

Eliminating  the velocity operator using
 the constraint converts the first equation
into a quadratic expression, which 
is  the quantum version of
the acceleration-dependent system of Lukierski et al.
\cite{LSZ1},
and can be obtained as a tricky non-relativistic limit
of Polyakov's ``particle with torsion''
\cite{HP1,Polyakov}.

The system (\ref{DL}) realizes  the ``exotic''
[i. e., two-fold centrally extended] planar Galilean
symmetry \cite
{exotic}. The commutator of the velocity operators is in
fact $[v_-,v_+]=2\kappa^{-1}$ where the
 real constant $\kappa=\theta m^2$ is
indeed the second central charge that
measures the extent Galilean boosts fail to commute \cite{
exotic}.
The Hamiltonian and the constraint weakly commute,
\begin{equation}
    \Big[h,\lambda_-\Big]=\frac{1}{m\theta}\lambda_-.
    \label{freeconsrel}
\end{equation}
The relation (\ref{freeconsrel}), which is the archetype of
``good'' behaviour,
guarantees  that the physical states do not mix
 with the unphysical ones during time evolution, i.e.,
the {\it consistency} of the system.
\goodbreak

The equations (\ref{DL}) only describe  free particles,
though, and coupling them to electromagnetic  field is not
entirely trivial.
The standard ``minimal coupling'' rule $\vp\to\vp-e\vA$,
simply inserted into both the Hamiltonian and the
constraint, yields in fact also some unwanted terms,
 see (\ref{inconsrel}) below.

We would also like to remind the reader to the analogous
difficulties
encountered in the relativistic case: the first-order,
Majorana-Dirac type anyon field equations
considered in \cite{JNany, Plany,MP1} as well as those put
forward
by Dirac \cite{Dirac}, only describe free particles.
Their coupling to external electromagnetic field  is
still an unsolved problem.

In this paper, we find a Hamiltonian and a constraint
which involve the electromagnetic field and such
that they {\it weakly commute}.

A general framework,
which includes both minimal and also nonminimal coupling,
is presented.
Special attention is paid to a critical case.
The quantum Landau problem and the associated field
equations are studied in detail.

Our paper is organized as follows. In Section 2
 the free classical symplectic model which
underlies our field-theoretical system (\ref{DL})
is  generalized so that it  accommodates
arbitrary background magnetic and electric
fields. The system we obtain
is described by two second class constraints,
 which,  at the critical
value of magnetic field, are transmuted into first
class constraints.

Section 3 is devoted to the
analysis of classical system in the critical case.
It is performed at the level
of the equations of motion,
proceeding from the generic case and
developing a Hamiltonian analysis.
For minimal coupling
the reduced Hamiltonian is given by the initial
scalar potential corrected by a  term, which is
quadratic in the electric field.

In Section 4 our analysis is extended to the quantum
case. First we identify the quantum Hamiltonian
and constraints, and then show that, when the
magnetic field is inhomogeneous, the
theory reveals a hidden nonlocal structure.

Then we analyse in detail the Landau problem in a constant
$B$-
field, for which theory is local,
and the generalization of the
 free anyon field equations (\ref{infLL}) are found.

A special section is devoted to the critical case, and its
relation to the generic (noncritical) Landau problem.

Our results are summarized in Section \ref{Conc}.

\section{Classical framework}\label{classical}

\subsection{The free symplectic model}

 The classical counterpart of the system (\ref{DL}) is
 given  by its symplectic structure
 and a pair of real second class constraints
\begin{eqnarray}
\omega=dp_i\wedge dx_i +\frac{1}{2}
\kappa\epsilon_{ij}
dv_i\wedge dv_j,
\label{symp0}
\\[6pt]
\lambda_i=p_i-mv_i\approx 0,\qquad
\{\lambda_i,\lambda_j\}=-\theta^{-1}\epsilon_{ij},
\label{lambda}
\end{eqnarray}
augmented with a  Hamiltonian,
presented in either of the equivalent forms
\begin{eqnarray}
    h=p_iv_i-\frac{1}{2} mv_i^2=
       \frac{1}{2}mv_i^2+v_i\lambda_i=
      \Di\frac{1}{2m}(p_i^2-\lambda_i^2).
   \label{h3}
\end{eqnarray}
The Lagrange multipliers ($v_i$) in the middle expression
 guarantee the conservation of the constraints,
$\dot{\lambda}_i=\{\lambda_i,h\}\approx 0$ \footnote{The
notation $\approx$ means
``on-shell'' i. e. after restriction to the constrained
surface.}.
The system has the conserved angular momentum
\begin{equation}
J=\epsilon_{ij}x_ip_j+\frac{1}{2}\theta m^2v_+v_-.
\label{Jfree}
\end{equation}

The second class constraints (\ref{lambda})
reduce the number of physical phase space degrees of
freedom from $6$ to $4$.
The gauge-invariant extension of the original coordinates
$x_i$,
\begin{equation}
X_i=x_i-\theta\epsilon_{ij}\lambda_j,
\label{X}
\end{equation}
and the momenta satisfy  the relations
$\{X_i,\lambda_j\}=\{p_i,\lambda_j\}=0$,
and can be therefore identified with dynamical variables
describing
the physical degrees of freedom (observables) of the system.
By (\ref{symp0}),
\begin{equation}
\{p_i,p_j\}=0,\quad
\{X_i,p_j\}=\delta_{ij},\quad
\{X_i,X_j\}=\theta\epsilon_{ij}.
\label{Xp0}
\end{equation}
Let us stress, in particular, that the new coordinates
$X_{i}$ are
non-commuting.
The angular momentum is presented equivalently as
\begin{equation}
J\approx \epsilon_{ij}X_ip_j +\frac{1}{2}\theta p_i^2.
\label{JXp}
\end{equation}

The Poisson bracket relations (\ref{Xp0})
imply that, when restricted to the surface of
second class constraints,  the classical system
is equivalent to the free exotic particle of \cite{DH}.

\subsection{Interactions}\label{interac}

Let us assume that the magnetic and electric
fields are static,
given by abelian vector and scalar potentials,  $A_i(\vx)$
and $V(\vx)$,
$B=\epsilon_{ij}\p_iA_j$, and $eE_i=-\p_iV$, respectively.
Let us consider the usual minimal coupling rule
\begin{equation}
    p_{i}\to P_{i}=p_{i}-eA_{i}(\vx).
    \label{pP}
\end{equation}
Inserting (\ref{pP}) into the Hamiltonian and the
constraints,
\begin{eqnarray}
h\to \widetilde{H}=
\vP\cdot
 \vv -\frac{1}{2}m
\, \vv{}\,{}^2+V(\vx)
\qquad\hbox{and}\qquad
\lambda_i\to
\widetilde{\Lambda}_i=P_i-mv_i\approx 0,
\label{Lam}
\end{eqnarray}
respectively, would yield additional terms in the Poisson
bracket,
\begin{equation}
    \Big\{\widetilde{H},\widetilde{\Lambda}_i\Big\}
    =\frac{1}{m\theta}\widetilde{\Lambda}_i+eBv_i+\p_iV.
     \label{inconsrel}
\end{equation}
violating the consistency relation (\ref{freeconsrel}).

Similarly,
$
\{P_i,\widetilde{\Lambda}_j\}= eB\epsilon_{ij}\neq0
$
so that the $P_i$ in (\ref{pP}) are not observable.

Below we correct these defects.
 On account of
\begin{equation}
\{\widetilde{\Lambda}_i,\widetilde{\Lambda}_j\}=
-\theta^{-1}(1-\beta)\epsilon_{ij}
\qquad\hbox{where}\qquad
\beta=\beta(\vx)=e\theta B(\vx),
\label{tLij}
\end{equation}
it is necessary to distinguish two cases.
Let us define the critical magnetic field by putting
\begin{equation}
B_{c}=\displaystyle\frac{1}{e\theta}.
\label{Bc}
\end{equation}

When $B\neq B_{c}$, the constraints (\ref{Lam}) are
second class, but for $B=B_c$ they turn into first
class. These cases should be analyzed separately.

We first consider the generic case $B\neq B_{c}$.
Let us first assume that we only have a magnetic field, and
try to generalize the free (kinetic) Hamiltonian in
(\ref{h3}) as
\begin{equation}
H_B=\frac{1}{2}mv_i^2+u_i\widetilde{\Lambda}_i.
\label{H0tL}
\end{equation}
Requiring that the constraints (\ref{Lam}) be
preserved,
$
\displaystyle\frac{d}{dt}\widetilde{\Lambda}_i\approx 0,
$
fixes the Lagrange multipliers as $u_i=(1-\beta)^{-1}v_i$,
and we get
\begin{equation}
H_B=\frac{1}{2}mv_i^2+v_i\Lambda_i,
\qquad\hbox{where}\qquad
\Lambda_i=\displaystyle\frac{1}{1-\beta}\,
\widetilde{\Lambda}_i\approx 0.
\label{Li}
\end{equation}
Then
$$
\{H_B,\Lambda_i\}=
\left(
\left(\displaystyle\frac{1}{m\theta}\delta_{jl}
+\frac{1}{2}v_j \epsilon_{lk}\partial_{k}
\right)(1-\beta(x))^{-1}\right)\Lambda_l\epsilon_{ji}
$$
weakly vanishes, as expected.

As long as $\beta\neq1$, the new  constraints $\Lambda_i$
in (\ref{Li}) are equivalent to the old ones in (\ref{Lam})
and indeed satisfy
\begin{equation}
\{\Lambda_i,\Lambda_j\}=
\left(-\displaystyle\frac{1}{\theta(1-\beta)}+
\Delta\right)\epsilon_{ij}
\approx
-\frac{1}{\theta(1-\beta)}\,
\epsilon_{ij},
\qquad
\Delta=\frac{1}{2}
\epsilon_{ij}\Lambda_i\partial_j
(1-\beta)^{-1}\approx0.
\label{LLD}
\end{equation}

The $\Lambda_i$ can also be presented in a form similar to
(\ref{lambda}),
\begin{equation}
\Lambda_i=\cP_i-mv_i\approx 0
\qquad\hbox{with}\qquad
\cP_i=\displaystyle\frac{1}{1-\beta}\,
(P_i-m\beta v_i).
\label{cPi}
\end{equation}
The new ``momenta'' $\cP_i$, unlike the $P_i$, {\it are}
observable,
$
\{\cP_i,\Lambda_j\}=\epsilon_{ij}\Delta\approx 0.
$
The Hamiltonian (\ref{Li}) has again equivalent forms,
namely
\begin{eqnarray}
H_B=\cP_iv_i-\frac{1}{2}mv_i^2
     =\displaystyle\frac{1}{2m}(\cP_i^2-\Lambda_i^2).
        \label{cHcP02}
\end{eqnarray}

The generalization of the free coordinate  $X_i$ in
(\ref{X}),
\begin{equation}
\cX_i=x_i-\theta\epsilon_{ij}\Lambda_j,
\label{cXx}
\end{equation}
is also observable,
$
\{\cX_i,\Lambda_j\}=\theta\Delta\delta_{ij}\approx 0.
$
Putting $\Omega= (1-\beta)^{-1}+\theta\Delta$,
\begin{eqnarray}
\Big\{\cX_i,\cX_j\Big\}=&
\theta\Omega\,\epsilon_{ij}\hfill
&\approx\displaystyle\frac{\theta}{1-\beta}\,
\epsilon_{ij},\label{cXcXcomrel}
\\[8pt]
\Big\{\cX_i,\cP_i\Big\}=&
\Omega\,\delta_{ij}\hfill
&\approx\displaystyle\frac{1}{1-\beta}\,\delta_{ij},
\label{cXcPcomrel}
\\[8pt]
\Big\{\cP_i,\cP_j\Big\}=&
\displaystyle\frac{\Omega-1}{\theta}\,\epsilon_{ij}\hfill
&\approx\displaystyle\frac{eB}{1-\beta}\epsilon_{ij}.
\label{cPcPcomrel}
\end{eqnarray}
The  angular
momentum  (\ref{JXp}) is now generalized to
\begin{equation}
J=\epsilon_{ij}\cX_i\cP_j
+\frac{1}{2}\theta\cP_i^2+\frac{1}{2}eB\cX_i^2.
\label{JcXP}
\end{equation}
It generates rotations of the observables $\cX_i$ and
$\cP_i$.

Having identified the observable variables which
correspond to the physical degrees of freedom,
now we extend the Hamiltonian in (\ref{Li})
by adding a scalar potential.

$\bullet$  Let us first consider
\begin{equation}
H=H_B+V(\cX),
\label{minham}
\end{equation}
for which the (\ref{freeconsrel})-type consistency condition
$
\Big\{H,\Lambda_i\Big\}\approx 0
$
holds.

Since $\cX_i\approx x_i$, the derivative
$\partial_i V(\cX)=-eE_i(\cX)$ is
($-e$ times) the electric field.
 (\ref{minham}) is viewed therefore
as a generalization of the minimally coupled Hamiltonian.
 It generates the equations of motion
\begin{eqnarray}
&m^*\dot{\cX}_i\hfill=\left(\big(
\cP_i+m\theta\epsilon_{ij}\partial_jV(\cX)\big)\,
\Omega
-\theta\Delta\Lambda_i
\right)(1-\beta)\hfill
\approx\cP_i-em\theta \epsilon_{ij}E_j,&
\label{vitesse1}
\\[8pt]
&m^*\dot{\cP}_i\hfill=\theta^{-1}\epsilon_{ij}\big(m^*\dot{
\cX}_j-
(1-\beta)\cP_j\big)\hfill
\approx
eB\epsilon_{ij}\cP_j+emE_i.&
\label{Lorentz1}
\end{eqnarray}
where
\begin{equation}
m^*=m(1-\beta)=m(1-e\theta B)
\end{equation}
is the effective mass.
The second equation from (\ref{Lorentz1})
on account of the first one can be presented
in the equivalent form
\begin{equation}
\dot{\cP}_i\approx
eB(\cX)\epsilon_{ij}\dot{\cX}_j+eE_i(\cX),
\label{PXEi}
\end{equation}

On-shell, our extended scheme reduces hence to
that of \cite{DH}.

Off the critical case, the variables
$\cX_{i}, \cP_{i}$ are observable and provide us with a
satisfactory description of the system in terms of
the constraints $\Lambda_{i}$ (\ref{cPi})
and the Hamiltonian $H$
(\ref{minham}).
The interaction can, however, also be
discussed in terms of the vector
\begin{equation}
Y_i=x_i+m\theta\epsilon_{ij}v_j,
\label{Y}
\end{equation}
whose use will be particularly convenient in the
critical case.
In terms of $\cX_i$ and $\cP_i$, it
can be presented equivalently as
\begin{equation}
Y_i=\cX_i+\theta\epsilon_{ij}\cP_j.
\label{Y*}
\end{equation}

$Y_{i}$ is, like $\cX_i$ and $\cP_i$,
observable, but unlike these, it {\it strongly}
commutes with the constraints,
\begin{equation}
    \{Y_i,\Lambda_j\}=0.
\end{equation}
$Y_i$ is decoupled from the coordinates $\cX_i$ and
satisfies
\begin{equation}
\{Y_i,\cX_j\}=0,\qquad
\{Y_i,\cP_j\}=\delta_{ij},
\label{YPX}
\end{equation}
\begin{equation}
\{Y_i,Y_j\}=-\theta\epsilon_{ij}.
\label{YY}
\end{equation}
In terms of  $\cP_i$ and $Y_i$,
(\ref{vitesse1})
takes the  form familiar from  point mechanics,
\begin{equation}
m\dot{Y}_i\approx \cP_i.
\label{PdotX}
\end{equation}

Using the decoupled, independent observables $\cX_i$
and $Y_i$,  the generator of rotations, (\ref{JcXP}),
can be represented in a quadratic normal form,
\begin{equation}
J=\frac{1}{2\theta}\left(Y_i^2-(1-\beta)\cX_i^2\right).
\label{JYX}
\end{equation}

$\bullet$ Another Hamiltonian can also
be considered now:
\begin{equation}
\check{H}=H_B+\check{V}(Y).
\label{nonmin}
\end{equation}
where, to avoid confusion with the previous case, we called
the
potential $\check{V}$.

It also satisfies the classical counterpart of
(\ref{freeconsrel}), $\{\check{H},\Lambda_i\}\approx 0$.
The expansion
$$
\check{V}(Y)=\check{V}(\cX)-\theta
e\epsilon_{ij}\check{E}_i(\cX)\cP_j+
\ldots
$$
allows us to infer that
$\partial_i\check{V}(Y)=-e{\check{E}}_i(Y)$
is the electric field only if $\check{V}(Y)$ is linear.
The  Hamiltonian (\ref{nonmin}) describes therefore a
particle
 with non-commuting coordinates and with non-minimal
 coupling.
The associated equation of motion can be written in the form
\begin{equation}
m^*\dot{\cX}_i\approx \cP_i,\qquad
m^*\dot{\cP}_i\approx
eB\epsilon_{ij}\cP_j+em^*\check{E}_i(Y),
\label{EMnm}
\end{equation}
cf. Eqs. (\ref{PdotX}) and (\ref{Lorentz1}),
respectively\footnote{The similarity between (\ref{EMnm})
and
(\ref{PdotX}) suggests  a kind of ``duality'' between the
two types
of couplings. Another generalization includes anomalous
coupling  \cite{anom}.}.

For a constant magnetic field $B=B_0\neq B_{c}$,
 the Lorentz force law (\ref{PXEi})
of the minimal coupling case
can  be presented in the equivalent form
\begin{equation}
\dot{\cX}^{gc}_i\approx \frac{1}{B_0}
\epsilon_{ij}E_j(\cX),
\label{Xgmin}
\end{equation}
where
\begin{equation}
{\cX}^{gc}_i:=\cX_i+\frac{1}{eB_0}
\epsilon_{ij}\cP_j.
\label{Xgc}
\end{equation}
In the nonminimal case we have instead,
using (\ref{EMnm}),
\begin{equation}
\dot {\cX}^{gc}_i\approx \frac{1}{B_0}
\epsilon_{ij}\check{E}_j(Y).
\label{Xgnm}
\end{equation}
${\cX}^{gc}_i$ can therefore be interpreted,
in both cases, as the guiding center coordinate
\cite{Ezawa}.

Compared to $Y_i$, ${\cX}^{gc}_i$ behaves in the
opposite way: it is decoupled from
$\cP_i$  (but not from $\cX_i$),
$\{\cX^{gc}_i,\cP_j\}\approx 0$ [that is behind
its evolution law (\ref{Xgmin}) or (\ref{Xgnm})],
and its brackets depend on the value of the constant
magnetic field,
\begin{equation}
\{\cX^{gc}_i,\cX^{gc}_j\}\approx
-\frac{1}{eB_0}\epsilon_{ij}.
\label{XXgc}
\end{equation}

Let us stress, however, that the definition
(\ref{Xgc}) of the guiding
center coordinate is restricted to
a homogeneous magnetic field.
Naively extended to the inhomogeneous case,
 $\cX^{gc}_i$ will perform  circular motion
with amplitude proportional to the gradient of $B$,
i.e. it will be a  guiding center coordinate
in zero order of $\partial_i B$.
The brackets (\ref{XXgc}) also will be corrected
by the term proportional to the gradient
of magnetic field.

We conclude this section with presenting the second-order
form of our equations of motion.

$\bullet$ In the minimal case
 (\ref{vitesse1}) -- (\ref{Lorentz1}) we have
\begin{equation}
\displaystyle\frac{d}{dt}
\big(
m^*\dot{\cX}_i+m\theta e\epsilon_{ij}E_j(\cX)\big)
\approx
eB(\cX)\epsilon_{ij}\dot{\cX}_j+
eE_i(\cX).
\label{dotcX}
\end{equation}
Expressed in terms of the coordinates $Y_i$, this reads
\begin{equation}
m^*\ddot{Y}_i
\approx
eB(\cX)\epsilon_{ij}
\dot{Y}_j+eE_i(\cX),
\label{dotX}
\end{equation}
where  the arguments of the magnetic and electric fields are
given by
$
\cX_i=Y_i-m\theta\epsilon_{ij}
\dot{Y}_j.
\label{cXX}
$

$\bullet$ In the non-minimal case (\ref{nonmin}) we have
instead
\begin{equation}
\displaystyle\frac{d}{dt}(m^*\dot{\cX}_i)
\approx
eB(\cX)\epsilon_{ij}\dot{\cX}_j+
e\check{E}_i(Y),
\label{dotcX*}
\end{equation}
where it is assumed that $Y_i=\cX_i+m^*\theta\epsilon_{ij}
\dot{\cX}_j$.
Equivalently, in terms of the $Y_i$,
\begin{equation}
m^*\displaystyle
\frac{d}{dt}(\dot{Y}_i-e\theta\epsilon_{ij}
\check{E}_j(Y))
\approx
eB(\cX)\epsilon_{ij}\dot{Y}_j+
e\check{E}_j(Y),
\label{nm2Y}
\end{equation}
where it is assumed that the argument of $B$
is expressed via the relation
$\cX_i=Y_i-m\theta\epsilon_{ij}\dot{Y}_j
-m\theta^2\check{E}_i(Y)$.

Note that for constant magnetic
 and electric fields, $B=B_0=const$, and
$E_i=\check{E}_i=E^0_i=const$,  respectively, the equations
(\ref{dotcX}), (\ref{dotX}), (\ref{dotcX*}) and (\ref{nm2Y})
take all the same form, namely
\begin{equation}
m^*\ddot{Z}_i=eB_0\epsilon_{ij}\dot{Z}_j+eE^0_i,
\label{Z}
\end{equation}
where $Z_i$ is either $\cX_i$ or $Y_i$.

\section{The critical case $B=B_{c}$}\label{critical}

For $B=B_{c}$ the constraints $\Lambda_{i}$
as well as the observables $\cP_i$ and $\cX_i$
are all ill-defined;
the variable $Y_i$ in (\ref{Y}) behaves in turn regularly.

The guiding center coordinates (\ref{Xgc}) are, a priori,
only defined off the
critical case. The divergences are readily seen to cancel
as $B_0\rightarrow B_c$, however, and
(\ref{Xgc}) becomes precisely $Y_i$, presented in the form
(\ref{Y*}).
The brackets (\ref{XXgc}) take
the form 
(\ref{YY}).
Thus, in the critical case,
$Y_{i}$ becomes the guiding center coordinate.

$\bullet$  For the minimal coupling, (\ref{minham})
at $B=B_c$, Eq. (\ref{vitesse1}) is reduced to
$\cP_i\approx
em\theta\epsilon_{ij}E_j$, and $Y_i$ becomes,
using Eq. (\ref{Y*}) and $\cX_{i}\approx x_i$,
\begin{equation}
Y_i\approx  x_i-\displaystyle\frac{m}{eB_c^2}E_i(\vx).
\label{guidcent}
\end{equation}
The equations of motion (\ref{dotX}) become now first order,
\begin{equation}
\dot{Y}_i\approx
\displaystyle\frac{1}{B_c}
\epsilon_{ij}E_j(\vx),
\label{dXcrit}
\end{equation}
cf. Eqn. (\ref{Xgmin}),
and we indeed recognize $Y_i$ as the familiar expression
of the guiding center in the Hall effect.

$\bullet$
Similarly in the non-minimal case (\ref{nonmin}),
at $B=B_c$ the first equation from  (\ref{EMnm}) gives
$\cP_i\approx 0$, and with $\cX_{i}\approx x_i$,
from Eqn. (\ref{Y*}) we find
\begin{equation}
Y_i\approx x_i.
\label{Yxnm}
\end{equation}
Then, on account of Eq. (\ref{dotcX*}) with $m^*=0$,
we find that
the {\it guiding center} follows the law
\begin{equation}
\dot{Y}_i\approx\displaystyle\frac{1}{B_c}
\epsilon_{ij}\check{E}_j(Y),
\label{YHall}
\end{equation}
cf. Eqn. (\ref{Xgnm}).

Both equations (\ref{dXcrit}) and (\ref{YHall})
are reminiscent of the {\it Hall law},
$\dot{Z}_i=\frac{1}{B_c}\epsilon_{ij}E^0_j$,
to which they both reduce if the
electric field is homogeneous.
In the general case, however, they are
slightly different~: on the one hand,
in (\ref{dXcrit}) the argument of electric field
is $x_i$, which is related to $Y_i$ via Eqn.
(\ref{guidcent}).
On the other hand, as noted above,
$\check{E}_i$ in (\ref{YHall})
is the electric field only
for $\check{E}_i=const$.


After these preliminary observations, we present a
Hamiltonian analysis of the critical case.

For $B=B_{c}$ we have $\beta=1$;
the constraints (\ref{Lam}) become
first class, and reduce therefore the dimension of the
physical phase subspace by $4$, rather than by $2$.
In all cases (critical or not),
$
\{Y_i,\tilde{\Lambda}_j\}=0.
$
In the critical case the noncommuting variables $Y_i$
represent the two independent phase space degrees of freedom
of the physical subspace. As  already said, the relation of
the initial coordinates $x_i$ to $Y_i$ depends on the choice
of the potential.

For the consistency of the theory, the conservation of the
constraints (\ref{Lam}) has to be checked.

$\bullet$ Let us first consider the minimally coupled system
 with
 $B=B_{c}$.
We seek again our Hamiltonian
 in the form
\begin{equation}
H=\frac{1}{2}mv_i^2
+u_i\tilde{\Lambda}_i+V(\vx),
\label{cHum}
\end{equation}

The first two terms here are as in (\ref{H0tL}), but the
argument of the potential has been changed from
$\cX_{i}$ [which is ill-defined for $B=B_{c}$] to
$x_{i}(\approx \cX_i)$.
The conservation of the constraints
$\tilde{\Lambda}_i\approx 0$
results in the gauge-fixing conditions
\begin{equation}
\chi_i=v_i+\theta\epsilon_{ij}
\partial_jV(\vx)\approx 0,
\qquad\hbox{i. e.}\qquad
P_{i}\approx m\epsilon_{ij}\frac{E_{j}(\vx)}{B_{c}}
\label{chivV}
\end{equation}
which played a r\^ole in the Hamiltonian reduction in \cite{
DH}.

The conservation of (\ref{chivV}) requires in turn,
\begin{equation}
M_{ij}u_j=v_i,
\qquad
M_{ij}=\delta_{ij}+m\theta^2\partial_i\partial_j V.
\label{uvV}
\end{equation}
When the matrix  $M_{ij}$ is non-singular, the constraints
(\ref{Lam})
and gauge conditions
(\ref{chivV}) provide us with four second class constraints;
then the  equations (\ref{uvV}) can be solved for the $u_i$.

Let us mention for completeness that,
for a repulsive oscillator potential
$$
V(x)=-\alpha\frac{eB}{2m\theta} x_i^2+\mu_i x_i+\nu,
$$
where $\alpha$, $\mu_i$ and $\nu$ are constants,
the matrix $M_{ij}$ in (\ref{uvV}) vanishes if $\alpha=1$.
Then (\ref{uvV}) gives new constraints $v_i\approx0$,
and we get $6$ second class constraints and reduction yields
a zero-dimensional phase subspace (i. e. a point) with fixed
values
$$
x_i=Y_i=m\theta^2\mu_i,
\quad
p_i=eA^c_i(x)_{x_i=m\theta^2\mu_i},
\quad
\epsilon_{ij}\partial_iA^c_j=B_c.
$$
This  can be understood as follows.
The equations of motion (\ref{dotcX})
with $V(\cX)=V(x)_{x_i=\cX_i}$ specified above, now read
\begin{equation}
(1-\beta)\ddot{\cX}_i+(\alpha-1)\frac{eB}{m}\epsilon_{ij}
\dot{\cX}_j
-\frac{\alpha}{m\theta}\,
\frac{eB}{m}\cX_i +\frac{1}{m}\mu_i=0.
\label{E2}
\end{equation}
When $B=B_{c}=(e\theta)^{-1}$
and $\alpha\neq 1$, we have the Hall-like  law
(\ref{dXcrit})
[whereas for
$\alpha=1$, eq. (\ref{E2})  reduces to
$\cX_i=m\theta^2\mu_i$].
When  $B\neq B_{crit}$ and $\alpha=1$, for
$\beta(1-\beta)<0$
and $\beta(1-\beta)>0$,
the system performs rotational resp. hyperbolic
motion around the  point $\cX_i=m\theta^2\mu_i$.
Hence, $\alpha=1$, $B=B_c$  corresponds to the
boundary that separates these two phases.

Returning to the generic case,
(\ref{chivV}) says that the variables $v_i$ are determined
by the Hall law,
\begin{equation}
v_{i}\approx\epsilon_{ij}\frac{E_{i}(\vx)}{B_c}.
\label{vE}
\end{equation}
The interpretation of
this relation requires some care, however~:
in the coupled case we consider here,
the variables $v_i$ do {\it not} represent
anymore the time
derivative of the original position.
Assuming that $M=\left(M_{ij}\right)$ is non-singular,
the ``velocity'' equation reads in fact
\begin{equation}
    \dot{x}_{i}=u_{i}=\left(M^{-1}\right)_{ij}v_{j}.
    \label{xdot}
\end{equation}
On the other hand,
$\dot{Y}_{i}=\{H,Y_{i}\}=v_i$,
which identifies  $v_{i}$ as the time derivative of the
guiding center coordinate $Y_{i}$; this latter
satisfies the Hall-like law, (\ref{dXcrit}).

On account of the relation
$(M^{-1})_{ij}=(\det\, M)^{-1}\epsilon_{ik}
\epsilon_{jl}M_{kl}$, equations (\ref{xdot})
with $v_i$ given by (\ref{vE})
can be presented in Hamiltonian form,
\begin{equation}
    \dot{x}_i=\{x_i,H_c\}
\qquad\hbox{with}\quad
H_c=V(\vx)+\frac{m}{2}
\theta^2(\partial_i V(\vx))^2
\label{cHcx}
\end{equation}
and
\begin{equation}
\{x_i,x_j\}=-\theta (\det\, M)^{-1}\epsilon_{ij},
\qquad
\det\, M =1+m\theta^2
\partial_i^2 V
+\frac{1}{2}(m\theta^2)^2\epsilon_{kl}\epsilon_{rs}(\partial
_k
\partial_r V)(\partial_l\partial_s V).
\label{xxM}
\end{equation}
The Hamiltonian (\ref{cHcx}) is indeed the reduction of
(\ref{cHum}) to the surface defined by
the second class constraints
(\ref{Lam}) and (\ref{chivV}),
with (\ref{xxM}) the corresponding Dirac brackets.
Note that in the  Landau problem
($V=0$), Eq. (\ref{guidcent}) is reduced to $Y_i\approx
x_i$,
and the brackets (\ref{xxM}) coincide with those of $Y_i$.

Alternatively, these coordinates $Y_i$ can  be used to
describe also the
system reduced to the surface (\ref{Lam}), (\ref{chivV}).
Their  brackets are $\{Y_i,Y_j\}=-\theta\epsilon_{ij}$
 cf. (\ref{YPX}), and the dynamics (\ref{dXcrit}) is
 generated by the Hamiltonian
\begin{equation}
H_c={\cal V}(Y)
\qquad\hbox{with}\qquad
{\cal V}(Y)=\left(V(\vx)+\frac{m}{2}
\theta^2(\partial_i V(\vx))^2\right)_{x_i=x_i(Y)}\,,
\label{cHc}
\end{equation}
where $x_i(Y)$
is given by (\ref{guidcent}).
This also explains the
advantage of quantizing the system in terms of the $Y_{i}$.
The reduced Hamiltonian $H_c$ extends the rule called ``
Peierls substitution'' \cite{DJT} to the non-commutative
case. The $V(Y)$ alone used in \cite{DH},
 obtained dropping the $\theta$-term,
 is only  correct for constant fields.

$\bullet$ The non-minimal  Hamiltonian is instead
\begin{equation}
\check{H}=\frac{1}{2}mv_i^2
+u_i\tilde{\Lambda}_i+\check{V}(Y).
\label{cHu}
\end{equation}
The conservation of the
first class constraints requires now
\begin{equation}
\chi_i=v_i\approx 0.
\label{chiv}
\end{equation}
The functions $\chi_i$ are such that
$\det||\{\varphi_a,\varphi_b\}||\neq 0$,
where $\varphi_a$, $a=1,2,3,4$, are
$\varphi_a=(\tilde{\Lambda}_i,\chi_j)$.
 (\ref{chiv}) is a gauge-fixing
 for the constraints (\ref{Lam}) \cite{GT};
the conservation of (\ref{chiv}) fixes furthermore the
Lagrange multipliers  as
$$
u_i=v_i+\theta\epsilon_{ij}\partial_j \check{V}(Y)\approx
\theta\epsilon_{ij}\partial_j\check{V}(Y).
$$
Reduced to the surface given by the  set of
second class
constraints (\ref{Lam}) and gauges (\ref{chiv}), the system
is described by the Hamiltonian
\begin{equation}
\check{H}_c=\check{V}(Y).
\label{cHcrit}
\end{equation}
It follows from  $\{Y_i,\tilde{\Lambda}_j\}=0$,
that the Dirac bracket
of the reduced phase space coordinates $Y_i=x_i$
coincides with their initial Poisson bracket (\ref{YY}).
The equation of motion of the
system (\ref{cHu}) is therefore (\ref{YHall}).
With hindsight to Eq. (\ref{JYX}), note that
in the critical case the generator of rotations is reduced
to
\begin{equation}
J=\frac{1}{2\theta} Y_i^2.
\label{JYXc}
\end{equation}

\section{Quantization and field equations}

Now we quantize the coupled system.
The different behaviour of the  constraints
$\widetilde{\Lambda}_i\approx 0$ and
$\Lambda_i\approx 0$ for $B\neq B_{c}$ and $B=B_{c}$,
respectively,
obliges us to distinguish between these two cases
also at the quantum level.
As we observed in Section \ref{interac}, the
$\widetilde{\Lambda}_i$ are  classically equivalent
to $\Lambda_i$ for $B\neq B_c$, but, unlike the latters,
the $\widetilde{\Lambda}_i$
are well defined also in critical case.
We start therefore, with the former constraints.

\subsection{The generic case $B\neq B_{c}$}

Let us start with the noncritical case. As in the free case,
we pass over to the conjugate complex linear combinations
\begin{equation}
\widetilde{\Lambda}_-
=\widetilde{\Lambda}_1-i\widetilde{\Lambda}_2\approx 0
\qquad\hbox{and}\qquad
\widetilde{\Lambda}_+
=\widetilde{\Lambda}_1+i\widetilde{\Lambda}_2\approx 0.
\end{equation}

The first combination here can be viewed
as a first class constraint and the
second one as a gauge condition for it.
Then, instead of quantizing  the system
with two second class constraints
which generate complicated, field-dependent Poisson-Dirac
brackets
on reduced phase space,
(\ref{cXcXcomrel})--(\ref{cPcPcomrel}),
we quantize it by the Gupta-Bleuler method.
This amounts to using the simple  symplectic structure (\ref
{symp0})
in total phase space,
and then selecting the physical quantum states by the
quantum constraint
\begin{equation}
\tilde{\Lambda}_-|\phi\rangle=0,\qquad\hbox{where}\qquad
\tilde{\Lambda}_-={P}_--m{v}_-.
\label{quL}
\end{equation}
As a result, we get a correspondence between the classical
and quantum descriptions
in  that, for any two physical states,
$
\langle\phi_1|\tilde{\Lambda}_\pm|\phi_2\rangle=0,
$
where
$
{\tilde{\Lambda}}_+
={\tilde{\Lambda}}_-^\dagger.
$

Any operator ${\cal O}$
which leaves the physical subspace (\ref{quL})
 invariant can be viewed as a quantum observable.
Therefore, it has to satisfy a relation of the form
$$
[\tilde\Lambda_-,{\cal O}]=(...)\tilde\Lambda_-
$$
with the operator $\Lambda_-$
appearing on the right, cf. (\ref{freeconsrel}).
This happens, in particular, for the operator
$Y_i=x_i+m\theta\epsilon_{ij}v_j$, which
strongly commutes with the constraint also
quantum-mechanically, and is, therefore, a quantum
observable.

We have to identify a quantum Hamiltonian, together with two
other independent  observables [see the classical equation
(\ref{Y})].
The  quantum counterpart of the `magnetic Hamiltonian'
$H_B$
is chosen to be the Hermitian analog of (\ref{H0tL}), namely
\begin{equation}
H_B=\frac{m}{2}v_+v_-+u_-^\dagger\tilde\Lambda_-
+\tilde\Lambda_+u_-.
\label{qH0}
\end{equation}

The operator-valued coefficient  $u_-$ is fixed here by
the observability requirement for (\ref{qH0})
as $u_-=\frac{1}{2}\cT v_-,$
where the operator $\cT$ [further discussed below]
is given formally by
\begin{equation}
\cT=\left(1+\frac{1}{2}\,
\frac{\theta}{1-\beta}\,
\tilde\Lambda_+\tilde\Lambda_-\right)^{-1}
\frac{1}{1-\beta}=
\frac{1}{1-\beta}
\left(
1+\frac{1}{2}\theta\tilde{\Lambda}_+
\tilde{\Lambda}_-\frac{1}{1-\beta}\right)^{-1}.
\label{qu-}
\end{equation}
The quantum analogs of the classical position
and momentum operators $\cX_\pm$ and  $\cP_\pm$ are
\begin{eqnarray}
\cX_+&=x_++i\theta\tilde{\Lambda}_+\cT,\qquad
\cX_-=\cX_-^\dagger\label{XQu},
\\[8pt]
\cP_+&=\tilde\Lambda_+\cT+mv_+,
\qquad \cP_-=\cP_+^\dagger,
\label{PQu}
\end{eqnarray}
cf. (\ref{cXx}), (\ref{cPi}).
Minimal and non-minimal coupling, respectively,
are obtained adding to $H_B$  the
 scalar potential with the $\cX_{i}$ resp. $Y_i$ in its
 argument.
 Decomposition of the second operator factor in (\ref{qu-})
into a formal infinite operator series shows, however, that,
due to the noncommutativity of $\tilde\Lambda_-$ and
$\beta(x)$,
 the theory is in general nonlocal in $x_i$
--- except for a homogeneous magnetic field,
discussed below.

\subsection{Constant magnetic field $B\neq B_c$}

Let us assume that the magnetic field is homogeneous
$B=const$, $B\neq B_c$.
Then  the operators $\tilde\Lambda_-$ and
$(1-\beta)^{-1}$ commute and the action of $\cT$
on  physical states 
 reduces to
multiplication by the constant $(1-\beta)^{-1}$. Thus
\begin{equation}
u_-=\frac{1}{2(1-\beta)}\,v_-,
\end{equation}
cf. Section \ref{classical}, and the
kinetic Hamiltonian is
\begin{equation}
H_B
= \frac{1}{2}\,(\cP_+v_- +
v_+\Lambda_-)=
\frac{1}{2m}(\cP_{+}\cP_{-}-\Lambda_+\Lambda_-)\, .
\label{quantBham}
\end{equation}
The observables (\ref{XQu}) and (\ref{PQu})
become now local operators
\begin{eqnarray}
\cX_+=x_++i\theta{\Lambda}_+,\qquad
\cX_-=\cX_-^\dagger,
\label{opXQu}
\\[6pt]
\cP_+=\Lambda_++mv_+=\frac{1}{1-\beta}(P_+-m\beta v_+),
\qquad \cP_-=\cP_+^\dagger,
\label{opPQu}
\end{eqnarray}
and
$\Lambda_\pm=(1-\beta)^{-1}\tilde{\Lambda}_\pm$,
cf. the classical relations
(\ref{Li}), (\ref{cPi}), and (\ref{cXx}).

Equation (\ref{quL}) means that the
physical states are coherent states, namely the
eigenstates of the velocity operator $v_-$ with
eigenvalue $P_-/m$.
The physical states, defined as solutions of (\ref{quL}),
are
\begin{equation}
|\phi\rangle_{phys}=\exp \left(\frac{1}{2}\theta mP_-v_+
\right)
\left(|0\rangle_v |\tilde{\phi}\rangle\right),
\label{statephys}
\end{equation}
where $|0\rangle_v$, $v_-|0\rangle_v=0$, is the vacuum state
of the Fock space generated by the velocity operators,
and $|\tilde{\phi}\rangle$ is a velocity-independent state
associated with other degrees of freedom
\footnote{Various  ``kets'' -- distinguished sometimes by
lower indices --
``live'' in different spaces.}.

The action of the observable operators on physical states
is reduced to
\begin{eqnarray}
&\cP_-\rightarrow P_-,\qquad
\cP_+\rightarrow
\displaystyle\frac{1}{1-\beta}P_+,\label{physP}\\
&
\cX_-\rightarrow x_-,\qquad
\cX_+\rightarrow x_++i
\displaystyle\frac{\theta}{1-\beta}P_+,
\qquad
Y_-\rightarrow x_-+i\theta P_-,\qquad
Y_+\rightarrow x_+,
&
\label{physX}
\end{eqnarray}
in the sense
\begin{equation}
\cP_-|\phi\rangle_{phys}=\exp \left(\frac{1}{2}\theta
mP_-v_+\right)\left(|0\rangle_v\,
P_-|\tilde{\phi}\rangle\right),
\end{equation}
etc., i.e. the operators on the right hand sides
act on $|\tilde{\phi}\rangle$.
Similarly, the action on  physical states
of the magnetic Hamiltonian (\ref{quantBham})
and of the quantum analog of the angular momentum
(\ref{JcXP}) reduce to
\begin{equation}
H_B\rightarrow \frac{1}{2m^*}P_+P_-
\label{Hred0}
\end{equation}
and
\begin{equation}
J\rightarrow \frac{i}{2}(x_+P_--x_-P_+)+\frac{1}{2}eBx_+x_-,
\label{Jred}
\end{equation}
respectively. On the subspace spanned by the
velocity-independent states $|\tilde{\phi}\rangle$,
let us define the operators
\begin{equation}
R_+=P_++ieBx_+,\qquad
R_-=P_--ieBx_-,
\label{RR}
\end{equation}
which correspond to
$\cP_++ieB\cX_+=ieB\cX^{gc}_+$ and
$\cP_- -ieB\cX_-=-ieB\cX^{gc}_-$,
acting in total space,
where $\cX^{gc}_i$ is the guiding center
coordinate (\ref{Xgc}). On account of the commutation
relations
\begin{equation}
[P_+,P_-]=2eB,\quad
[P_-,x_+]=[P_+,x_-]=-2i,\quad
[P_+,x_+]=[P_-,x_-]=0,
\label{PPxx}
\end{equation}
the operators (\ref{RR})
commute with $P_+$, $P_-$ and satisfy the relation
\begin{equation}
[R_+,R_-]=-2eB.
\label{RRB}
\end{equation}
They reduce the angular momentum operator (\ref{Jred})
to normal Hermitian form,
\begin{equation}
J\rightarrow \frac{1}{2eB}\, (R_+R_--P_-P_+).
\label{JRP}
\end{equation}

The  commutation relations
(\ref{PPxx}) and (\ref{RRB}) depend on the sign of $eB$.
We have to distinguish therefore two cases. For both signs,
we have two independent sets of
creation-annihilation oscillator operators $a^\pm$ and
$b^\pm$.
The cast is sign-dependent, though~:

$\bullet$ For $eB<0$,
\begin{equation}
a^-=\frac{1}{\sqrt{2|eB|}}\,P_-,\quad
a^+=\frac{1}{\sqrt{2|eB|}}\,P_+,\quad
b^-=\frac{1}{\sqrt{2|eB|}}\,R_+,\quad
b^+=\frac{1}{\sqrt{2|eB|}}\,R_-,
\label{ab-}
\end{equation}
satisfy
$[a^-,a^+]=[b^-,b^+]=1$, $[a^\pm,b^\pm]=0$.
In their terms, the angular momentum operator
(\ref{JRP}) takes the canonical quadratic  form
\begin{equation}
J\rightarrow \, a^+a^--b^+b^-.
\label{Jab-}
\end{equation}

$\bullet$
For $eB>0$ we have, instead of (\ref{ab-}),
\begin{equation}
a^-=\frac{1}{\sqrt{2eB}}\,P_+,\quad
a^+=\frac{1}{\sqrt{2eB}}\,P_-,\quad
b^-=\frac{1}{\sqrt{2eB}}\,R_-,\quad
b^+=\frac{1}{\sqrt{2eB}}\,R_+,
\label{ab+}
\end{equation}
and the angular momentum operator reads
\begin{equation}
J\rightarrow \, b^+b^--a^+a^-.
\label{Jab+}
\end{equation}

Let us now assume that we have a purely magnetic field.
Hence $H=H_{B}$, and angular momentum is conserved.
Consider the physical states (\ref{statephys}) with
$$
|\tilde{\phi}\rangle=|n\rangle_a|l\rangle_b,
\qquad\hbox{where}\qquad
a^+a^-|n\rangle_a =n|n\rangle_a,
\quad
b^+b^-|l\rangle_b =l|l\rangle_b,
$$
$n,l=0,1,\ldots$, i.e. consider the states of the form
\begin{equation}
|n,l):=e^{\frac{1}{2}\theta mP_-v_+}
|0\rangle_v |n\rangle_a|l\rangle_b.
\label{NBisstates}
\end{equation}

$\bullet$
For $eB<0$, the (\ref{NBisstates}) are eigenstates of
operators
$H=H_B$ and $J$ with eigenvalues
\begin{equation}
E_N=\frac{|eB|}{m^*}N,\quad N=n=0,1,\dots
\qquad\hbox{and}\qquad
j=n-l=N,N-1,\ldots,
\label{E1}
\end{equation}
respectively.
The energy spectrum is therefore discrete and nonnegative;
each Landau level is infinitely degenerate
in the angular momentum, which takes integer values
bounded from above.

$\bullet$ For $eB>0$, the roles of the operators $P_+$ and
$P_-$
 as creation and annihilation operators are interchanged,
 and instead of (\ref{E1}) we have
\begin{equation}
E_N=\frac{eB}{m^*}(N+1),\quad N=n=0,1,\dots,
\qquad\hbox{and}\qquad
j=l-n=-N,-N+1,\ldots.
\label{EN2}
\end{equation}
Here we should distinguish two further ``phases''\footnote{
Two phases were discussed also in the
context of another noncommutative
quantum mechanical model in
\cite{BNS}, which is
related to the model \cite{DH}
by a time rescaling, $t\to (m/m^*)t$, supplemented with
(nonunitary) change of variables.}.

$\bullet\bullet$ For $0<eB<\theta^{-1}$,
the energy spectrum is discrete and positive.
The Landau levels are infinitely degenerate
in $j$, which takes an infinite number of integer values and
is
bounded from below.

$\bullet\bullet$ For $eB>\theta^{-1}$, we have the same
degeneration
of Landau levels, but the effective mass
$m^*$ becomes negative. In order
to make the theory  well defined and to eliminate
the negative-energy,  unbounded-from-below spectrum,
we change the sign of the evolution parameter,
$t\rightarrow -t$.  This is
equivalent to changing the sign of the
Hamiltonian operator. The energy spectrum is
given therefore by  (\ref{EN2}) but with $m^*$ changed into
$|m^*|$.
Below we shall see, however, that even with such a change
of time evolution parameter,
the cases $0<eB<\theta^{-1}$ and $eB>\theta^{-1}$
correspond to the two essentially different phases.

One could attempt to restore the spectrum symmetry around
$B=0$,
by changing the quantum ordering. Replacing indeed
the quantum Hamiltonian (\ref{quantBham}) by
\begin{equation}
H_B^s=\frac{1}{4}(\cP_+v_-+v_-\cP_+)+\frac{1}{2}v_+\Lambda_-
=\,
\frac{1}{4m}(\cP_+\cP_-+\cP_-\cP_+)-\frac{1}{2m}\Lambda_+
\Lambda_-
\label{HBsym}
\end{equation}
i. e.
$
H_B^s\rightarrow\frac{1}{4m^*}(P_+P_-+P_-P_+)
$
on the physical states (\ref{statephys})
yields, instead of (\ref{E1}) and (\ref{EN2}),
 the spectrum
\begin{equation}
E_N=\frac{|eB|}{m^*}\left(N+\frac{1}{2}\right),
\qquad N=0,1,\dots
\label{Esym}
\end{equation}
which looks to be symmetric w.r.t.  $B=0$.
The asymmetry between $B<0$ and
$B>0$  is still present, however,
since it is hidden in
 the asymmetric behaviour of the effective mass.
To second order in $B$, the spectrum is indeed
$|B|(1+e\theta B)(N+1/2)$.

Parity invariance is hence violated by
 planar noncommutativity, and this is revealed by coupling
to a magnetic field.

Let us now investigate the question of
normalizability of the states (\ref{NBisstates}).
Let us define the `normalized' velocity creation-
annihilation oscillator operators
\begin{equation}
c^\pm=\sqrt{\frac{\theta m^2}{2}}\,v_\pm,\qquad
[c^-,c^+]=1,
\end{equation}
and the corresponding Fock states $|k\rangle_v$,
$c^+c^-|k\rangle_v=k|k\rangle_v$,
$k=0,1,\ldots$.
Decomposing the exponential factor into Taylor series, we
find that for
$eB>0$ the states (\ref{NBisstates})
are given by  an infinite  superposition
\begin{equation}
|n,l)=\sum_{k=0}^{\infty} \beta^{k/2}
\sqrt{C^{n+k}_k}\,
|k\rangle_v |n+k\rangle_a|l\rangle_b,\qquad
C^{n+k}_k=\frac{(n+k)!}{k!\,n!}
\label{Nsuper+}
\end{equation}
that satisfy the relation
\begin{equation}
( n',l'|n,l)=\delta_{n'n}\delta_{l'l}
\sum_{k=0}^{\infty}
\beta^k C^{n+k}_k.
\label{Njscalar}
\end{equation}

$\bullet$
For $0<eB<\theta^{-1}$, we have $\beta<1$ and the series
in (\ref{Njscalar}) converges.
The orthonormal Landau states are
\begin{equation}
|n,l\rangle = {\cal N}^{-1/2}|n,l),\qquad
{\cal N}={\cal N}(\beta,n)=\frac{1}{n!}\,
\frac{d^{n}}{d\beta^{n}}
\left(\frac{1}{1-\beta}\right).
\label{NJnorm+}
\end{equation}

$\bullet$
For $eB>\theta^{-1}$, $\beta>1$ and the series
in (\ref{Njscalar}) diverges.
In this case the energy eigenstates
(\ref{Nsuper+}) are scattering-like, unbounded states.

Hence, the two cases
$0<eB<\theta^{-1}$ and $eB>\theta^{-1}$
correspond to two, essentially different phases.

$\bullet$
For $eB<0$, operator $P_-$ is in fact the annihilation
operator
$a^-$ [see Eq. (\ref{ab-})]. As a result, (\ref{NBisstates})
is a superposition of $(n+1)$ states,
\begin{equation}
|n,l)=\sum_{k=0}^{n}|\beta|^{k/2}\sqrt{C^{n}_k}\,
|k\rangle_v|n-k\rangle_a|l\rangle_b.
\label{Nsuper-}
\end{equation}
The corresponding orthormal states are given by
\begin{equation}
|n,l\rangle = {\cal N}^{-1/2}|n,l),\qquad
{\cal N}={\cal N}(\beta,n)=(1-\beta)^{-n/2}.
\label{NJnorm-}
\end{equation}

In conclusion, the structure of the states
(\ref{NBisstates})
is essentially different for $eB<0$ and $eB>0$:
in the first case it is a normalizable superposition
of $(n+1)$ velocity Fock states;
in the second case it is an infinite superposition
of all  velocity-Fock states, which is
normalizable (bound state) for $0<eB<\theta^{-1}$,
and is not normalizable (scattering-like state)
for $eB>\theta^{-1}$.
In this sense the noncomutative Landau problem
for $B\neq B_c$
has three, essentially different phases
(plus the critical case phase separating
the phases $0<eB<\theta^{-1}$ and $eB>\theta^{-1}$,
see below).

The hidden finite resp. infinite
dimensional structures of the physical
states in the $eB<0$ and
$eB>0$ phases described here are reminiscent of the
finite resp. infinite-dimensional
representations of the $sl(2,R)$ algebra,
associated with universal
Majorana-Dirac-like equations for
usual integer/half-integer resp. anyonic fields
in $2+1$ dimensions \cite{MP1},
see below.

Note that such a  hidden structure
would be absent if we quantized the system
on the reduced phase space given by the
second class constraints.
This  would eliminate effectively the velocity degrees of
freedom
($v_\pm$), and yield the system in Ref. \cite{DH},
described only by the variables $\cX_i$ and $\cP_i$, and
would have
 symplectic structure
(\ref{cXcXcomrel}), (\ref{cXcPcomrel}), (\ref{cPcPcomrel}).
Then, a superposition of the states of the form
$|\tilde{\phi}\rangle=|n\rangle_a|l\rangle_b$
would describe the quantum states of the system.
In such a quantization scheme
it would be impossible to reveal the difference between
the energy eigenstates for
$eB<\theta^{-1}$ and $eB>\theta^{-1}$
in the sense of their normalizability.
This is consistent with an observation
\cite{PR} which says that different quantization methods
can produce physically inequivalent results.

To conclude this section, we
generalize the free field equations (\ref{infLL})
\begin{equation}
     \left\{\begin{array}{cccc}
     \left(i\p_{t}+eB
     \displaystyle\frac{k}{m^*}\right)\phi_{k}
     &+&\sqrt{ \displaystyle\frac{k+1}{2\theta}}\,
     \displaystyle\frac{P_+}{m^*}
     \,\phi_{k+1}
     &=0,
	\\[18pt]
	P_{-}\,\phi_{k}
	&+&\sqrt{\displaystyle\frac{2(k+1)}{\theta}}\,
	{\phi}_{k+1}
	&=0.
     \end{array}\right.
     \label{BinfLL}
\end{equation}
Each of these equations is just the component form of
the Schr\"odinger equation
$i\partial_t|\phi\rangle=H_B|\phi\rangle$
and constraint equation (\ref{quL}), respectively,
where $H_B$ is taken in the first, linear-in-$\cP$
form in Eq.~(\ref{quantBham}),
and the field components are defined as
$\phi_k(x,t)=(-1)^k({}_v\langle k|\langle
x|\phi\rangle)$.

Eliminating the $(k+1)^{th}$ component using the lower
equation
 we find that each  component satisfies  an independent
Schr\"odinger-Pauli type equation, namely
\begin{equation}
    i\p_{t}\phi_{k}=H_{k}\phi_{k},
    \qquad
    H_{k}=H_{0}-\frac{eB}{m^*}k,\qquad
    H_0=\displaystyle\frac{1}{2m^*}P_+P_-.
    \label{compBSch}
\end{equation}

 $H_{0}$ is, in particular, the restriction of the
Landau Hamiltonian $H_B$ to the velocity vacuum sector,
cf. Eq. (\ref{Hred0}).
For the higher components, the Hamiltonian $H_k$
looks formally like that for a charged particle
with planar spin $s=\frac{1}{2}k$
and with gyromagnetic ratio $g=2$.
Our system is a spinless, however~:
it is described just by
one independent field component $\phi_0$,
whose internal angular momentum (spin) is
zero \cite{HP2}. It can not
be treated as a scalar particle, however, see
 the last section below. The tower
of higher components is generated  by  $\phi_0$
according to
\begin{equation}
\phi_k=(-1)^k\left(
\displaystyle\frac{\theta}{2k!}\right)^{k/2}
 P_-^k\,
 \phi_0.
\label{phin}
\end{equation}
$\phi_0$ corresponds to the wave function
of the velocity-independent state $|\tilde{\phi}\rangle$
from the physical state (\ref{statephys}),
$\phi_0=\langle x|\tilde{\phi}\rangle$.
In general, it describes an
arbitrary superposition of the Landau states
$|n,l\rangle$.
In the particular case
$|\tilde{\phi}\rangle=|n,l\rangle$,
the corresponding fields $\phi^{n,l}_k(x,t)$
are the stationary states
of the Schr\"odinger-Pauli equation
(\ref{compBSch}).
In accordance  with Eqs.
(\ref{Nsuper+})
the tower of components
$\phi^{n,l}_k$
is infinite for $eB>0$, while according to
Eq. (\ref{Nsuper-}),
for $eB<0$ it contains a finite number of
nonzero components
$\phi^{n,l}_k$,
$k=0,1,...,n$.
In this respect, the structure of our field equations
is  {\it formally}  similar to that of the
infinite-component, relativistic Majorana-Dirac like
equations for (2+1)D anyons, and also to the finite-
component
equations for usual spin $s=\frac{1}{2}n$ fields \cite{MP1}.

\section{The critical quantum case}\label{cqc}

In the critical case the constraints $\widetilde{\Lambda}_i$
in (\ref{Lam}) are  first class.
Since the observables $Y_i$ strongly commute with the
$\widetilde{\Lambda}_i$,
the Hamiltonian can be taken, consistently with Section 3,
to be a function of the $Y_i$,
$H_c={\cal H}(Y)$. ${\cal H}(Y)$
is given by Eq. (\ref{cHc}) and  (\ref{cHcrit})
for  minimal or nonminimal coupling, respectively.
Here we assume that the observables
are realized in terms of the total (initial) phase space
variables
$Y_i=x_i+m\theta\epsilon_{ij}v_j$,
see
Eq. (\ref{Y}), while the constraints (\ref{Lam})
should be treated as equations specifying the
classical physical subspace.
 Then, following Dirac, condition (\ref{quL})
has to be supplemented at the quantum level with
\begin{equation}
\tilde{\Lambda}_+|\phi\rangle=0,\qquad
\tilde{\Lambda}_+=P_+-mv_+.
\label{L+}
\end{equation}
Using the solution (\ref{statephys}) of Eq. (\ref{quL}),
we find that equation (\ref{L+}) is reduced to
\begin{equation}
P_+|\tilde{\phi}\rangle=0.
\label{P+phi}
\end{equation}
The operators $P_+$ and $P_-$
satisfy here the commutation relation
$[P_+,P_-]=2\theta^{-1}$,
i.e. they act as
annihilation and creation operators.
Let us choose the gauge
\begin{equation}
A_+=A_1+iA_2=\displaystyle\frac{i}{2e\theta}\, x_+,\qquad
A_-=A_1-iA_2=-\displaystyle\frac{i}{2e\theta}\, x_-.
\label{gA1}
\end{equation}
Hence,
\begin{equation}
P_+=-2i\frac{\partial}{\partial x_-}-\frac{i}{2\theta}x_+,
\qquad
P_-=-2i\frac{\partial}{\partial x_+}+\frac{i}{2\theta}x_-.
\end{equation}
Then the solution of Eq. (\ref{P+phi}) is  given by
\begin{equation}
\tilde{\phi}(x_+,x_-)=\langle x_+,x_-|\tilde{\phi}\rangle=
\exp \left(-\frac{1}{4\theta}x_+x_-\right)\, f(x_+).
\label{Laughlin}
\end{equation}
where $f(x_+)$ is an arbitrary holomorphic function.
 Thus, we recover the wave functions proposed by Laughlin
to describe the ground states in the
Fractional Quantum Hall Effect \cite{Laugh, QHE}.
The action of the operators
$Y_\pm$ on these functions is reduced to
\begin{equation}
Y_+
\tilde{\phi}(x_+,x_-)
=e^{-\frac{1}{4\theta}x_+x_-}\, x_+
f(x_+),\qquad
Y_-\tilde{\phi}(x_+,x_-)
=e^{-\frac{1}{4\theta}x_+x_-}\,
2\theta\frac{d}{dx_+}f(x_+),
\label{YLaugh}
\end{equation}
cf. (\ref{physX}).

Due to the noncommutativity of the operators $Y_-$ and
$Y_+$,
the construction of the quantum Hamiltonian  ${\cal H}(Y)$
requires to choose an ordering.

Let us clarify now how such a quantum picture
is related to the generic  Landau problem
with a noncritical magnetic field.
First we note that, just like classically,
the critical case corresponds to the boundary between
two different phases, namely
$0<eB<\theta^{-1}$ and $eB>\theta^{-1}$, respectively.
In accordance with
Eqs. (\ref{physP}), (\ref{physX}),
the $Y_i$ are (unlike $\cP_+$ and $\cX_+$)
 well defined at $eB=\theta^{-1}$:
$$
Y_-\rightarrow i\sqrt{2\theta}\, b^-,
\qquad
Y_+\rightarrow -i\sqrt{2\theta}\, (b^+-a^-).
$$

Since at $eB=\theta^{-1}$ the second
constraint equation is reduced to Eq. (\ref{P+phi}),
we find that the physical states
 now belong to the lowest Landau level (LLL),
\begin{equation}
|0,l)=e^{\frac{1}{2}\theta mP_-v_+}
|0\rangle_v|0\rangle_a|l\rangle_b.
\label{Nstates0}
\end{equation}
By Eq. (\ref{Njscalar}),
such states are not normalizable.
We can `renormalize' the
theory by first projecting,
in the phase $0<eB<\theta^{-1}$, to the lowest Landau
level, and then taking the limit $\beta\rightarrow 1$.
The projection of an observable ${\cal O}$ to the lowest
Landau level is
\begin{equation}
({\cal O})_0=\Pi_0 {\cal O}\Pi_0
\qquad\hbox{where}\qquad
\Pi_0=\sum_{l=0}^\infty |0,l\rangle\langle 0, l|
\end{equation}
[for $N=n=0$,
$j=l$, see Eq. (\ref{EN2})].
Taking into account the definitions (\ref{RR}) and
(\ref{ab+}),
 we find that
\begin{equation}
({\cP}_\pm)_0=0,\quad
({\cX}_-)_0=({Y}_-)_0
=
i\sqrt{\frac{2}{eB}}\,b^-\,
\Pi_0,\quad
({\cX}_+)_0=({Y}_+)_0=-i\sqrt{\frac{2}{eB}}\,b^+\,
\Pi_0.
\label{ProjectionLLL}
\end{equation}
Therefore, when restricted to the lowest Landau level,
the  coordinates $\cX_i$ act identically as the $Y_i$ do.
Letting $\beta\rightarrow 1$ yields that, in the critical
case,
the action of these operators is reduced to
\begin{equation}
({\cX}_-)_0=({Y}_-)_0
\longrightarrow i\sqrt{2\theta}\,b^-
\qquad
({\cX}_+)_0=({Y}_+)_0\longrightarrow
-i\sqrt{2\theta}\,b^+,
\label{XYcrit}
\end{equation}
cf.  Eq. (\ref{YLaugh}),
where, on the right hand sides we assume that the
operators act on LLL  states.
These operators are the only independent
nontrivial operators of the system put
into a critical magnetic field,
and, in accordance with Eqns. (\ref{RR})
and (\ref{ab+}), are the
guiding center coordinate (\ref{Xgc})
reduced to the LLL states.
Note  that the relations (\ref{XYcrit})
correspond to the classical brackets (\ref{xxM})
taken for $V=0$ and
with  $x_i\approx Y_i$, valid
generically for nonminimal coupling with
$B=B_c$ (see Section 3).

The normalized states
(\ref{NJnorm+})
corresponding to the lowest Landau level
for $0<eB<\theta^{-1}$
are
$$
|0,l\rangle=\left((1-\beta)^{1/2}
\sum_{k=0}^\infty \beta^{k/2}
|k\rangle_v |k\rangle_a\right)|l\rangle_b.
$$
Taking the limit $\beta\rightarrow 1$ and
factorizing
the fixed state vector in the bracket 
yields the reduced quantum space of
those LLL states  $|l\rangle_b$.

The Laughlin wave functions (\ref{Laughlin})
are the states $|l\rangle_b$ written in
holomorphic representation,
in which the operators (\ref{XYcrit})
are reduced to (\ref{YLaugh}).

\section{Discussion and concluding remarks }\label{Conc}

The main result of this paper is a consistent scheme to
couple
the first-order, Majorana-Dirac type equations that describe
a non-relativistic anyon to an electromagnetic field.
(An alternative framework is presented in \cite{Stich}.)
Let us stress that the analogous results in the relativistic
setting are still
missing\footnote{
For relativistic models of anyons see
\cite{Polyakov,JNany,Plany,MP1,relany}.}.

Our framework here is analogous to the usual electromagnetic
coupling of a Dirac particle in terms of the initial
coordinates $x_i$, subject to Zitterbewegung. That in
\cite{DH}
corresponds in turn to the
introduction of electromagnetic
interaction using the Zitterbewegung-free Foldy-Wouthuysen
coordinates.

The Dirac equation for a spin-$1/2$ particle
can be treated as a constraint which removes the
timelike spin degrees of freedom (which otherwise would
produce negative norm states), and generates
the mass shell, Klein-Gordon equation.
The structure behind both of these properties
is the local supersupersymmetry of a Dirac
particle \cite{SUSYDirac}.
 Modification of a free Dirac equation according to the
minimal coupling prescription  produces, via
local supersymmetry, the correct form of
the mass shell equation.

For our system,  the constraint
makes, in the free case, the spectrum bounded from below by
freezing its
spin degrees of freedom.
So, it plays the role similar to that of the (first-order)
Dirac equation,
while the Schr\"odinger equation
is analogous to the (quadratic) mass-shell condition.
We have, however, no local supersymmetry here.
To switch on interaction
we use instead  weak commutativity as guiding principle.
This  fixes the
form of the Hamiltonian to be consistent with the
constraint,
modified  according to the minimal coupling prescription.

When the magnetic field is non-constant (inhomogeneous),
our quantum theory becomes non-local. This
property is actually inherited from relativistic anyons
from which our  theory can be derived \cite{HP2}. It is
indeed reminiscent of the
nonlocality of anyon fields within the framework of the
Chern-Simons U(1) gauge field construction,
with associated half-infinite
nonobservable `string', \cite{W,Sem,Ban}.

Viewing the critical case also as a ``phase'', we can say
that, in  the non-commutative Landau problem, our system
exhibits {\it
four} different phases.
These ``phases'' may be viewed as analogous to
various energy sectors of the Kepler problem,

$\bullet$ $E<0$\qquad ($eB<\theta^{-1}$),

$\bullet$ $E=0$\qquad ($eB=\theta^{-1}$),

$\bullet$ $E>0$\qquad ($eB>\theta^{-1}$)

\noindent
with its changing
$SO(4)$, $E(3)$ and  $SO(3,1)$ symmetry.

The phases on each side of $B=0$ exhibit
 a spectral asymmetry with respect
to the sign of magnetic field, and an essentially
different structure of the states:
for $eB<0$,  the angular momentum takes an
infinite number of values, $j=N,N-1,\ldots, -\infty$.
The field associated with a Landau state of level
$N$ has ($N+1$)-components.
On the other hand, for $0<eB<\theta^{-1}$,
the  field is infinite-component,
with the angular momentum taking $j=-N,-N+1,\ldots,+\infty$,
for a level-$N$ Landau state.
In both  phases the corresponding quantum mechanical
states are normalizable and represent bound states.

In the phase $eB >\theta^{-1}$,
for a Landau state of level $N$, the angular momentum
can take an infinite half-bounded set of integer values,
with, for any fixed value of $j$, an associated  infinite-
component field.
The corresponding quantum states are non-normalizable
scattering states.

In the critical case $eB\theta=1$,
our framework generalized to the non-commutative
context  the Peierl's substitution \cite{DJT} and Laughlin's
framework for the Fractional Quantum Hall Effect \cite{QHE}.

In the free case ($\cP_i=p_i$), our system
describes a spinless particle.
It cannot be interpreted, however,
as an ``ordinary''  scalar particle. In the rest frame
system ($\vp=0$)  it is described  effectively
by the velocity vacuum state  (see Eqn. (\ref{statephys}))
and by associated one field component only (constant in this
frame).
In a boosted frame, however, an infinite tower
of velocity Fock states contribute, coherently,  to the
physical state;   an infinite tower
of field components (\ref{phin}) is ``brought to life''.
Thus, the internal spin degrees
of freedom associated with velocity operators
(see Eqn. (\ref{Jfree})) are here frozen, rather then
removed as in the model \cite{DH}.  The
corresponding velocity Fock states $|k\rangle_v$ with
$k>0$
can be compared with a Dirac sea of the negative
energy states.

When we switch on a magnetic
field, the `frozen sea' `melts'.
For $eB<0$, in the state corresponding
to the $N^{th}$ Landau level, the first $N$ exited velocity
Fock states contribute to the state.
In contrast, for $eB>0$,  the entire, infinite set
of velocity Fock states is ``brought to life'', just like
in the free
case for $\vp\neq 0$.
Taking also into account the normalizability properties
of the physical states, the phases with
$eB<0$, $0<eB<\theta^{-1}$, $eB>\theta^{-1}$ and
$eB=\theta=1$ are somewhat reminiscent resp.
to a partially `melted', a `liquid',
a `gaseous' and `boiling' phases.

The observed spectral asymmetry
of the system is reminiscent of the phase transition between
exact (for $eB<0$) and spontaneously broken
(for $eB>0$) supersymmetry, see \cite{Wit}.
An analogous asymmetry has been observed earlier
in   noncommutative Chern-Simons theories, namely between
 self-dual and anti-self-dual solution, see \cite{Bak}.

In the noncritical case,
we  only analysed in detail the purely magnetic Landau
problem.
Adding a potential term $V(\cX)$ (or
$\check{V}(Y)$), the analog of first equation in
(\ref{BinfLL}) would involve other field components,
due to the presence of velocity operators
in operators $\cX_i$ and $Y_i$,
see Eqns. (\ref{opXQu}), (\ref{Y}).
 The second equation in (\ref{BinfLL}), which controls the
 number of nontrivial field components  would not change,
 however.
These latter behave, therefore
exactly in the same way as in the pure Landau problem.


\vskip 0.4cm\noindent
{\bf Acknowledgements}.
MP is indebted to the {\it Laboratoire de Math\'ematiques et
de Physique Th\'eorique} of Tours University
for hospitality extended to him.
The  partial support by the
FONDECYT, Chile (Grants 1010073 and 7010073)
is acknowledged.


\end{document}